# Current states of a doubly connected superconductor with film bridges

A. V. Krevsun, L. V. Gnezdilova, V. P. Koverya, and S. I. Bondarenko

*B.I. Verkin Institute for Low Temperature Physics and Engineering of the National Academy of Sciences of Ukraine, 47 Lenin Ave., Kharkov 61103, Ukraine*

Transport current distribution in the branches of a doubly connected superconductor in the form of a thin-film high-inductance circuit with two bridges of different width in the branches was measured. For the transport current lower than the sum of critical currents of the bridges, its distribution was found to exhibit an anomalous behavior upon reaching the critical current of one of the bridges. For a fixed value of the transport current through the circuit higher than the sum of critical currents of the bridges, low-frequency continuous harmonic voltage self-oscillations together with synchronous current self-oscillations appeared in the circuit branches. The mechanism responsible for the onset of the self-oscillations is discussed.

## Introduction

Current states in doubly connected circuits with one or two superconducting film bridges in their branches are at the heart of the most popular superconducting electronic devices.[1,2] In particular, a constant current through the circuit which consists of a superconducting branch with a bridge and a normal branch generates high-frequency voltage self-oscillations in the circuit.[3,4] If a dc quantum interferometer with two bridges is included in the superconducting branch of a high-inductance loop, the voltage self-oscillations become dependent on external magnetic field.[5] In the case of a doubly connected superconductor (DCS) in the form of a closed superconducting circuit with superconducting branches comprising a bridge contact or pressure point contact, which acts as a "weak" (in regard to the value of critical current) link, in one or both branches, a constant current through the circuit may cause low-frequency current self-oscillations in its branches.[6,7] It has also been found that the characteristics of the self-oscillations depend on the type of the weak link in the branches.

The purpose of this paper is to study the current states in a previously unstudied type of a high-inductance DCS with film bridges of different width in its branches at transport currents both lower or higher than the sum of the critical currents of the bridges.

## Experimental

The schematics of a film DCS-circuit containing two bridges is shown in Fig. 1. The inset shows the geometry of the bridges, which differ in width $w$, but have the same length $l$.

The film thickness was 120 nm, the length of the bridges $l \approx 6$ $\mu$m, and the widths of the bridges $w_1$ and $w_2$ were 15 and 25 $\mu$m, respectively. The film was deposited by thermal evaporation of indium-tin alloy (50% In–50% Sn) from a tantalum boat at the rate of approximately 10 nm/s and a pressure below $10^{-6}$ Torr on a glass-ceramics substrate (0.5 mm thick, the substrate temperature $T = 20\,°C$). The critical temperature of the deposited film $T_c \approx 5$ K. The required circuit geometry was obtained by photolithography using cleanroom equipment (Babcock). The photolithographic pattern was transferred to the film by chemical etching in a 5% aqueous solution of hydrochloric acid. The bridges were made by scribing the film with a diamond pyramid using a microhardness measurement instrument PMT-3. Electrical connections to the circuit contact pads were made with a 0.07 mm diameter copper wire using indium solder. The calculated inductance of the circuit was $L \approx 10^{-8}$ H. This inductance is considered high as compared with the allowable inductance of SQUID circuits (less than $10^{-9}$ H at 4.2 K). To determine the current states of the DCS, a constant transport current $I_t$ was injected from a current source into the circuit through the contacts as shown in Fig. 1 (the distance between the current terminals $I_t$ was about 1 mm). At the same time, the magnetic field $H_I$ generated by the currents in the circuit branches and the voltage $V$ in the circuit were measured. The magnetic field was measured using a fluxgate (FG) magnetometer with sensitivity to a homogeneous magnetic field of about $10^{-5}$ Oe. The longitudinal axis of the magnetometer was oriented perpendicular to the plane of the circuit and the distance between the surface and the extremity of the fluxgate probe was about 1 mm. The voltage on the circuit in resistive state of the bridges was measured with a photoelectric microvoltmeter F116. The studies of current states in the DCS were carried out with the DCS circuit placed in helium vapor to eliminate the effect of liquid helium boiling on the surface of the bridges, which is known to cause random changes in the circuit voltage.[8] This was achieved by placing the circuit in a brass cup the bottom of which was immersed in liquid helium. The top edge of the cup was above the liquid helium level. The circuit temperature was determined through the temperature of helium vapor used as a heat-exchange gas and was measured using a calibrated semiconductor thermometer located in the proximity of the circuit. To eliminate the influence of the electromagnetic excitation of the fluxgate sensor on the film circuit, the sensor was shielded by a copper cup. Schematics of the cryogenic insert containing the circuit and the fluxgate is shown in Fig. 2. The cryostat containing the cryogenic insert was equipped with a two-layer permalloy shield to protect the circuit and the fluxgate from external electromagnetic fields.

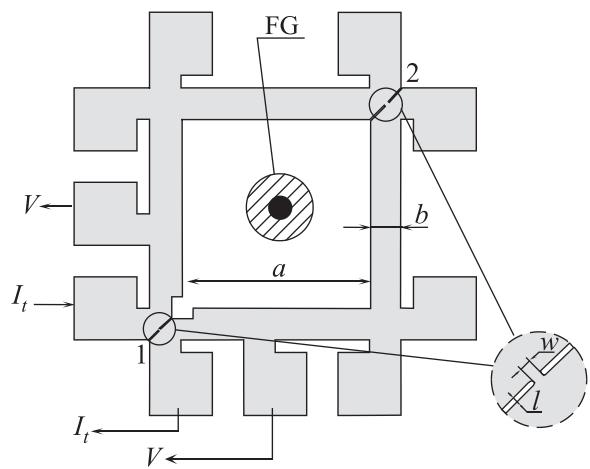

FIG. 1. Schematics of the film DCS-circuit with two bridges (1, 2). In the inset: $l$ and $w$ are the length and width of the bridge, FG is the schematic depiction of the double-rod fluxgate magnetometer with the rod axis oriented perpendicular to the image plane. The inner opening size $a = 4$ mm and the width of the branches $b = 0.5$ mm.

**Results and Discussions**

Fig. 3 shows the current-voltage characteristics (CVC) of the circuit with bridges at 4.6 K. The CVC hysteresis, which was observed when the transport current was first increased above the critical current of the circuit ($I_{c,t} = 43$ mA) and then subsequently decreased, indicates heating of the bridges, which causes their transition to normal state.[8] In this case, the onset of the circuit resistance was accompanied by harmonic voltage self-oscillations (SO$_V$) with a low frequency of about 2 Hz appearing on the circuit. The modulation depth of the SO$_V$ is 0.75 mV, i.e., about 5% of the mean value of the voltage on the circuit.

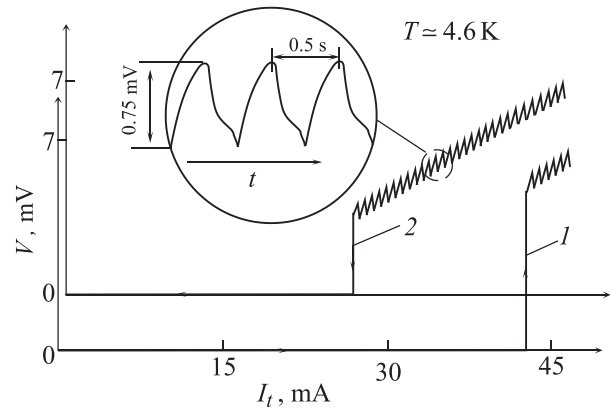

FIG. 3. Current-voltage characteristics of the DCS circuit with two bridges measured in helium vapor at $T_w = 4.6$ K using a linear current sweep: with increasing (*1*) and decreasing (*2*) current. The current-voltage characteristics (*1*) and (*2*) are offset along the ordinate axis for clarity. The inset shows a zoom of the data-recorder log displaying the voltage self-oscillations at a fixed value of the transport current through the circuit in the resistive region of the current-voltage characteristic.

Along with the voltage oscillations, the oscillations of magnetic field $H_I$ at the same frequency were registered in the circuit using the FG, hence, indicating current self-oscillations (SO$_I$). These oscillations occurred synchronously with the SO$_V$. Fig. 4 shows the ramping of the magnetic fields $H_I$ produced by the current in the DCS circuit, which was observed upon increasing the transport current $I_t$ through the circuit. This ramping exhibits two characteristic features: a change in the slope of the dependence $H(I_t)$ at $I_t = 17$ mA, and the emergence of SO$_V$ immediately after a sharp decrease in the field $H_I$ at $I_t = I_{c,t} = 43$ mA due to the resistance appearing in the branches of the DCS circuit.

First, we consider the origin of the SO$_V$ and SO$_I$. It should be noted that voltage self-oscillations are absent in the CVCs of the individual bridges of the circuit (Fig. 5). These characteristics were obtained after completion of the studies of current states in DCS by cutting the film circuit and hence destroying the doubly-linked geometry. This implies that the appearance of the SO$_V$ is characteristic for a doubly connected structure only.

The mechanism of occurrence of the SO$_I$ and the associated SO$_V$ can be explained in a similar way to what we did in regard to the current oscillations in the DCS circuit branches with two resistive pressure point contacts (PPCs).[7] In the circuit with PPCs, the self-oscillations arise from a periodic quantum change in resistivity of the micro interferometers forming the PPC structure. This occurs due to the magnetic field of the transport current in the branches of the circuit. In contrast to this, as will be shown below, in a thin film circuit with bridges the change in their resistance occurs due to periodic current overheating of one of the bridges. The bridges used in the experiments can be called "long," i.e., such that their length is much greater than the coherence length of the superconducting film (which does not exceed 100 Å). The theory of the transition to resistive (normal) state for single long and short film bridges placed in liquid helium or vacuum has been most fully developed by Skocpol, Beasley, and Tinkham (SBT theory).[8] The analysis based on this theory gives reason to believe that for the heat dissipation in the resistive bridge equal to about $10^{-4}$ W, which is typical for our experimental situation, their resistivity in helium vapor corresponds to the normal state. In the SBT theory, the emergence of normal resistance in such bridges is associated with the occurrence of the so-called hot spots in bridges upon reaching the critical current. The size of such spot in a bridge, the normal region inside it and near its edges, can change with increasing the transport current. Thus, the resistance of such bridges is parametric, i.e., current dependent. What is new as compared with the situation of the SBT theory is the parallel connection of two bridges which are different in critical current and resistance. The theory explaining the emergence of critical state in such superconducting structures is absent at present. Below we propose a qualitative explanation of the phenomena occurring in this case.

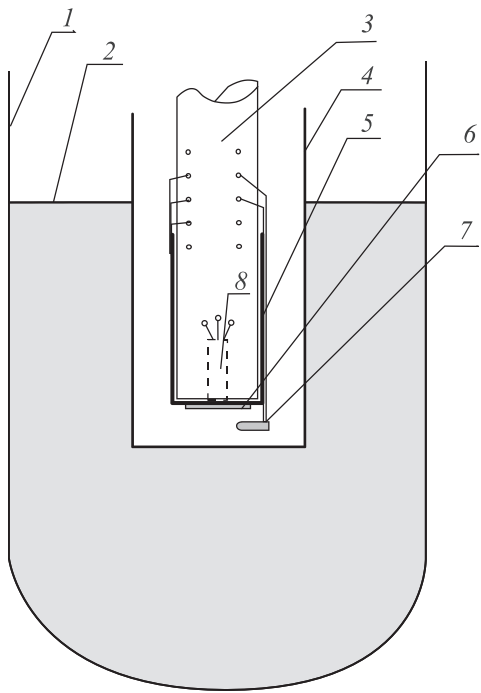

FIG. 2. Schematic design of the cryogenic insert for investigating the characteristics of the DCS circuit in helium vapor: cryostat (*1*), liquid helium level (*2*), cryogenic insert (*3*), brass cup (*4*), copper cup (*5*), sample (*6*), thermometer (*7*), and fluxgate (*8*).

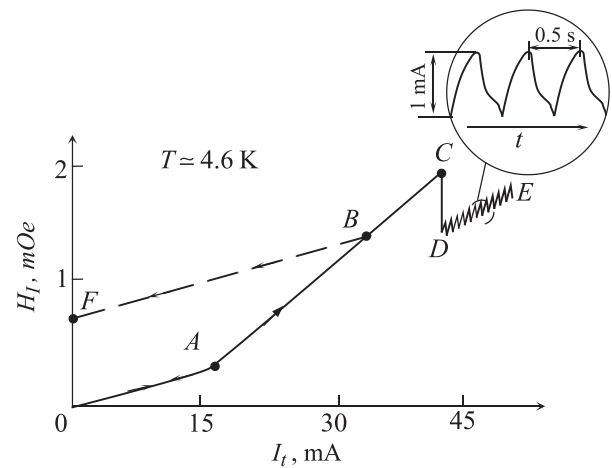

FIG. 4. Ramping of the magnetic field $H_1$ generated by the current in the DCS circuit upon linearly increasing the transport current $I_t$. The characteristic points on the curve $H_I(I_t)$ are denoted by letters *A, B, C, D, E, F*.

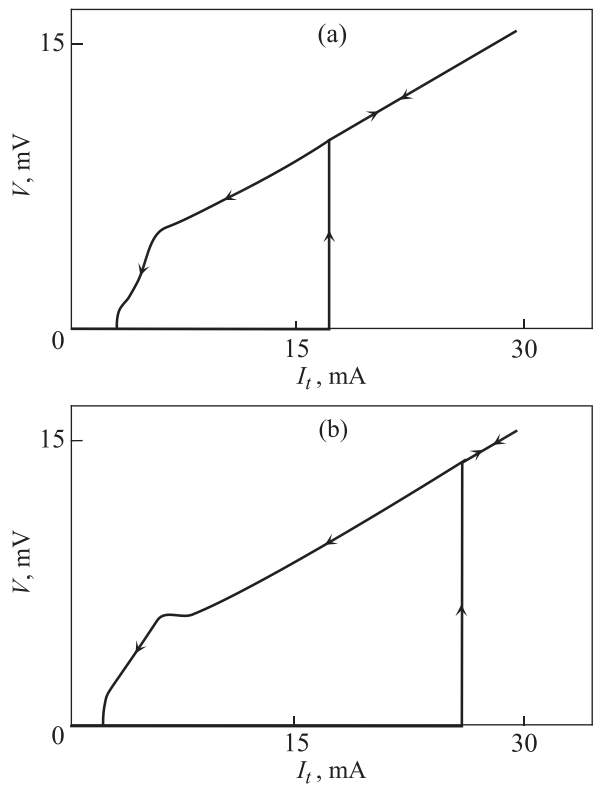

FIG. 5. The current-voltage characteristics of bridges 1 (a) and 2 (b), measured at $T_w = 4.6$ K in helium vapor (after cutting the DCS circuit).

In the case of the circuit with the bridges which have different critical currents, at $I_t \approx I_{c,t}$ the bridges exhibit different resistances $R_1^*$ and $R_2^*$, which depend on the magnitude of the currents $I_1$ and $I_2$ in the branches. Initially, when the critical current of the circuit is reached, the resistance $R_1^*$ of the first (smaller) bridge exceeds the resistance $R_2^*$ of the second (larger) one, and the total resistance of the circuit is

$$R_{01}^* = R_1^* R_2^* / R_1^* + R_2^*. \qquad (1)$$

At a given current $I_t$, the current $I_2$ exceeds $I_1$, and hence the heat dissipation $P_2 = I_2^2 R_2^*$ in the second bridge is higher than the respective dissipation ($P_1$) in the first. Upon further heating of the bridges in a gas environment (after a certain time $\Delta t_1$), due to the temperature dependence of their resistances, the increase in the resistance of the second bridge $\Delta R_2^*$ can become larger than that of the first one, $\Delta R_1^*$. Two consequences occur as the result of this process. First, the net resistance of the circuit increases up to

$$R_{02}^* = (R_1^* + \Delta R_1^*)(R_2^* + \Delta R_2^*)/(R_1^* + \Delta R_1^*) + (R_2^* + \Delta R_2^*). \qquad (2)$$

Second, a redistribution of the current between the branches of the circuit can occur, when some fraction of the current switches from the branch containing the second bridge with an increased resistance to the branch containing the first bridge. This will lead to an additional heating there and subsequent increase in the resistance of the first bridge and, hence, a new redistribution of the current. The first phenomenon can cause the experimentally observed short-term (during the time $\Delta t_1$) increase in voltage in the circuit, while the second reduces the voltage to its initial value with a subsequent increase. Thus, such a process can be repeated periodically, which qualitatively explains the observed voltage self-oscillations $SO_V$ in the circuit.

Let us now explain the frequency of the $SO_V$. We denote the amplitude of the observed voltage self-oscillations in the circuit as $\Delta V$ and the amplitude of the observed current self-oscillations as $\Delta I$. They correspond to the increment of the circuit resistance by $\Delta R^* = \Delta V / \Delta I$. Periodic heating and cooling of the circuit bridges occurs due to short-term variations of the transport current $\Delta I$ flowing through them. To estimate the period of these processes, we first determine the heat $Q$ released in the bridge due to the current $\Delta I$,

$$Q = \Delta I^2 \Delta R^* \Delta t_1, \qquad (3)$$

where $\Delta t_1$ is the experimentally observed time of voltage increase in the periodic dependence $V(t)$. According to the proposed model of self-oscillation, after one of the bridges has been overheated and the current redistribution between the branches has occurred, the cooling of the bridge begins. Let us estimate using a simple one-dimensional heat equation the dissipation time of this heat from the bridge into the film

$$\lambda S \Delta T / \Delta x = Q / \Delta t_2, \qquad (4)$$

where $\lambda$ is the thermal conductivity of the film, $\Delta x$ is the approximate distance along the film at which the temperature is reduced to its initial value (we assumed this distance to be equal to the length of the bridge), and $S$ is the cross section of the bridge.

Combining Eqs. (3) and (4) and substituting the following values: $\lambda \approx$ at 200 W/(m·K), $\Delta T = \Delta T_c - \Delta T_w = 0.4$ K, $\Delta I \approx 1$ mA, $\Delta V = 0.75$ mV, and $\Delta R^* \approx 0.75$ Ohm, we obtain $\Delta t_2 \approx 0.1$ s. Thus, the expected period of the oscillation is $t = \Delta t_1 + \Delta t_2 \approx 0.35$ s, which is close to the experimental value (0.5 s).

The provided estimate for the self-oscillations period supports the hypothesis about the role of thermal processes in the mechanism of voltage self-oscillation in a DCS circuit with the bridges the resistance of which depends parametrically on the current flowing through them.

Let us now consider the origin of the peculiarities in the dependence $H_I(I_t)$. According to the Laue law,[9] the transport current $I_t$ is distributed between the circuit branches until it reaches the critical current of the weakest bridge. Under the assumption of small mutual inductance between the branches of the circuit, this law has the form

$$I_1 = I_t(L_2/L); \quad I_2 = I_t(L_1/L), \qquad (5)$$

where $L_1$ and $L_2$ are the inductances of the first short branch with a narrow bridge and the second branch, respectively.

The calculated inductance[10] of the short branch does not exceed $5 \cdot 10^{-10}$ H. Given that $L = L_1 + L_2$ and $L_2/L_1 \gg 1$, it follows from Eq. (5) that for small values of $I_t$, the current in the first branch is significantly higher than the current in the second one. Increasing the current $I_t$ leads to a current increase in both branches, so the critical current $I_{c1}$ of the narrower (smaller) bridge in the short branch will be reached first. Similar to the DCS circuits with two PPCs,[7] a new non-resistive critical state is formed in the circuit with two bridges. In this state, the superconductivity of the circuit is preserved, but the current in the short branch $I_{t1}$ cannot exceed $I_{c1}$ and remains constant upon further increase in $I_t$ ($I_{t1} = I_{c1} = $ const). The superconducting current $I_t$ can only increase by increasing the current through the second branch with the larger inductance $L_2$ and the wider bridge. The onset of the first non-resistive state in the circuit manifests itself as an inflection in the dependence $H_I(I_t)$, i.e., change in its slope. Since $L_2/L_1 \gg 1$, the value $I_t$ which corresponds to the inflection in $H_I(I_t)$ is close to the critical current of the narrow bridge $I_{c1}$ and can be used to determine $I_{c1}$ without breaking the doubly connected geometry of the circuit.

The magnetic states of the superconducting circuit before and after the first current critical state are different as well. At $I_t < I_{c1}$ the magnetic flux through the circuit created by the currents in the branches remains constant and equal to zero. Increasing or decreasing $I_t$ causes a corresponding increase or decrease of the registered local field $H_I$ above the circuit, while no hysteresis occurs in this region of the dependence $H_I(I_t)$. After the first critical state of the superconducting circuit is reached, the magnetic flux generated by the current $I_2$ through the second branch is no longer compensated by the magnetic flux $\Phi_1$ generated by the current $I_1$ through the first branch. The net magnetic flux through the circuit $\Phi$ becomes non-zero. Thus, in accordance with the law of magnetic flux conservation in a superconducting circuit, decreasing the current $I_t$ from the values $I_t > I_{c1}$ down to zero results in freezing the flux. In contrast to freezing current and the corresponding magnetic flux in a DCS circuit with two PPCs,[7] the magnetic flux frozen in the circuit with two film bridges is not a discrete function of the transport current. The flux freezing leads to a hysteretic dependence $H_I(I_t)$ (see, for example, the dotted line B–F in Fig. 4).

When the current $I_t$ is increased up to $I_{c1} + I_{c2}$, there is a second, resistive, critical current state of the circuit, the characteristic features of which were discussed above.

## Conclusions

The conducted studies of the current states in a high-inductance DCS device with two "long" (length is much greater than the coherence length of the superconductor) In-Sn film bridges revealed that some of their characteristic features are identical to the states in a DCS circuit with two PPCs, while some others are typical only for a DCS circuit with bridges. The similarities include the presence of two critical current states (where the first is non-resistive and the second is resistive), the ability to freeze the transport current and its magnetic flux after the transport current exceeding the first critical current of the circuit is switched off, and the presence of current self-oscillations in the branches of the circuit after its transition into the second, resistive, state. The features typical only for the DCS with bridges are the following: the transport current in the branch with a higher inductance, as well as the frozen current and the flux in the DCS circuit exhibit linear (rather than quantized) dependence on the transport current through the DCS device, the current self-oscillations ($SO_I$) in the resistive state of the DCS circuit bridges exist for any given value of the transport current, and not only at some discrete values of it, and considerably large (about a millivolt) low-frequency voltage oscillations ($SO_V$) in the normal region of the CVC.

The obtained results show that the common features in current characteristics of these two types of DCS circuits originate from the asymmetry in the critical currents and inductances of their branches. The differences in the current states in these two types of circuits are associated with the specific properties of weak regions in the DCS branches. While in the case of DCS with PPCs the major role is played by macroscopic quantum phenomena (quantization of current and magnetic flux), in the case of DCS with relatively "long" (compared to the coherence length) bridges the main role is played by the phenomena that became classical, such as the Laue law, the resistance of the superconducting bridges in critical state, and their heat exchange with the surrounding environment. In particular, the appearance of $SO_V$ favors the hypothesis of the thermal nature of the parametric changes in the resistance of the bridges, which depend on the periodic increase in the transport current in one branch accompanied by a simultaneous transport current decrease in another branch.

In conclusion, the authors thank A. Omelyanchuk, A. G. Sivakov, and A. L. Solovyov for valuable comments during the discussion of the results.

[1] B. B. Schwartz and S. Foner, *Superconductor Applications: SQUIDs and Machines* (Frances Bitter National Magnet Lab., M.I.T. Cambridge, Massachusetts, Plenum Press, N.Y., 1977).

[2] W. P. Jolly, *Cryoelectronics* (The English Universities Press Ltd., London, 1972).

[3] H. Frohlich, H. Koch, W. Vodel, D. Wachter, and O. Frauenberger, Wissenschaftliche Zeitschrift, Der Friedrich–Schiller–Universitat, Jena, Heft 1/2, 1973.

[4] K. Enpuku, T. Kisu, and K. Yoshida, IEEE Trans. Magn. **27**, 3058 (1991).

[5] M. Muck and C. Heiden, IEEE Trans. Magn. **25**, 1151 (1989).

[6] S. I. Bondarenko, V. P. Koverya, A. V. Krevsun, N. M. Levchenko, and A. A. Shablo, Fiz. Nizk. Temp. **36**, 202 (2010) [Low Temp. Phys. **36**, 159 (2010)].

[7] V. P. Koverya, A. V. Krevsun, S. I. Bondarenko, and N. M. Levchenko, Fiz. Nizk. Temp. **38**, 44 (2010) [Low Temp. Phys. **38**, 35 (2012)].

[8] W. J. Skocpol, M. R. Beasley, and M. Tinkham, J. Appl. Phys. **45**, 4054 (1974).

[9] M. von Laue, *Theorie der Supraleitung* (Springer, Berlin, 1949).

[10] P. L. Kalantarov and L. A. Tseitlin, *Inductance Calculations* (Energoatomizdat, Leningrad, 1986) [in Russian].